\documentclass[aps,prl,preprint,groupedaddress,english]{revtex4-2}
\usepackage[caption=false,font=footnotesize,position=top]{subfig}
\usepackage{amsmath}
\DeclareMathOperator{\e}{e}
\DeclareMathOperator{\erfc}{erfc}
\usepackage{amssymb}
\usepackage[T1]{fontenc}% optional T1 font encoding
\usepackage[english]{babel}

\usepackage{pgfplots,pgf}
	\usetikzlibrary{shadows}
	\usetikzlibrary{plotmarks}
	\usetikzlibrary{spy,calc,positioning,shapes}
	\pgfplotsset{compat=newest}
	\pgfplotsset{plot coordinates/math parser=false}
	\newlength\figureheight
	\newlength\figurewidth

	\usepackage[per-mode=symbol]{siunitx}
	\DeclareSIUnit[]\bd
{Bd}
	\DeclareSIUnit[]\baud
{Bd}
	\DeclareSIUnit[]\photon
{ph}
	\DeclareSIUnit[]\symbol
{sym}
	\DeclareSIUnit[]\sym
{sym}
	\DeclareSIUnit[]\bit
{bit}
	\DeclareSIUnit[]\belm
{Bm}
	\DeclareSIUnit[]\voltpp
{V_{pp}}
	\DeclareSIUnit[]\sample
{Sa}
	\DeclareSIUnit[]\snu
{SNU}
	\DeclareSIUnit[]\ppm
{ppm}
	\DeclareSIUnit[]\sps
{sps}
\begin{document}
%\preprint{}

\title{Continuous-Variable Quantum Key Distribution with a Real Local Oscillator and without Auxiliary Signals}


\author{Sebastian Kleis, Max Rueckmann, Christian G. Schaeffer}
\email[]{kleis@hsu-hh.de}
%\thanks{}
\affiliation{Helmut-Schmidt-Universit\"{a}t Hamburg}


\date{\today}

\begin{abstract}
Continuous-variable quantum key distribution (CV-QKD) is realized with coherent detection and is therefore very suitable for a cost-efficient implementation.
The major challenge in CV-QKD is mitigation of laser phase noise at a signal to noise ratio of much less than \SI{0}{\deci\bel}. So far, this has been achieved with a remote local oscillator or with auxiliary signals.
For the first time, we experimentally demonstrate that CV-QKD can be performed with a real local oscillator and without auxiliary signals which is achieved by applying Machine Learning methods.
It is shown that, with the most established discrete modulation protocol, the experimental system works down to a quantum channel signal to noise ratio of \SI{-19.1}{\deci\bel}. The performance of the experimental system allows CV-QKD at a key rate of \SI{9.2}{\mega\bit\per\second}
over a fiber distance of \SI{26}{\kilo\meter}. After remote local oscillator and auxiliary signal aided CV-QKD, this could mark a starting point for a third generation of CV-QKD systems that are even more attractive for a wide implementation because they are almost identical to standard coherent systems.
% improvement of sensitivity?
\end{abstract}

%\keywords{}

\maketitle

\section{Introduction}
With quantum computers threatening to break the security of today's cryptosystems, the field of quantum communications attracts increasing attention. In this field, continuous variable quantum key distribution (CV-QKD) is very attractive for a practical implementation because it is based on coherent detection and promises to require only commercial off-the-shelf components.
However, one crucial difference to classical optical communications
is that the received power level is usually less than \SI{1}{\photon\per\bit} for a discrete modulation \cite{leverrier2011}, translating to a signal to noise ratio (SNR) of less than \SI{0}{\deci\bel}.
%In this sub-photon-level regime, phase tracking based on the received optical signal becomes much more challenging.
Additionally, residual phase noise contributes to the excess noise \cite{kleis2017}, which is the critical performance parameter for CV-QKD \cite{grosshans2003}. The combination of ultra low SNR and required accuracy makes carrier phase recovery the major challenge in practical CV-QKD.
The first approach to deal with this challenge was to avoid phase noise by using a remote local oscillator (LO) \cite{grosshans2003,lodewyck2005,qi2007,jouguet2013}. The disadvantage is that the remote LO compromises the security and limits the achievable distance \cite{jouguet2013a}.
Therefore, it is highly preferable that CV-QKD systems work with a real LO generated by a separate laser source at the receiver site (Bob). Since 2015, various real LO systems have been proposed and experimentally demonstrated \cite{kleis2015,huang2015,qi2015,soh2015,schrenk2016,marie2017,schrenk2017,comandar2017,laudenbach2017,wang2018}. 
In all the proposed systems, carrier phase estimation is based on auxiliary signals, also called pilots, that are generated at the transmitter site (Alice).
%On the one hand using pilot tones seems to be a viable approach to enable the use of a real LO.
However, the pilots occupy additional bandwidth and significantly increase the complexity of the system. For example, the system in \cite{laudenbach2017} uses only one polarization for the quantum signal and the orthogonal one for a pilot tone. In \cite{kleis2015}, two pilot tones are multiplexed with the quantum signal in the frequency domain and occupy more bandwidth than the quantum signal itself. In both cases, the spectral efficiency could at least be doubled without pilots.
%Additionally, the pilot tones and the quantum signal in \cite{kleis2015} are detected by the same balanced receiver which could have an impact on the receiver calibration.
Another important issue is that pilot tones are not included in the current security proofs for CV-QKD \cite{fossier2009,leverrier2011}.
%Their impact on the security of CV-QKD systems has not been studied yet.
Despite all these issues, no real LO system without pilot tones has been reported up to now.
Regarding the large number of pilot-based systems proposed, it appears as if pilot tones were indispensable in CV-QKD systems with a real LO. However, no convincing argument or experimental demonstration about the necessity of pilots has been made yet.
The present article addresses the question of whether or not designing a CV-QKD system without pilot tones is possible with current fiber-optic technology.
%It is a difficult task to give an answer to this just by theoretical analysis and simulations, mainly because the highly complex behavior of laser phase noise.
%Therefore, 
Our approach is to design and experimentally investigate such a CV-QKD system. In this, the quantum signal is discrete phase modulated with an order of $M=2$ or $M=4$. The latter corresponds to the most established CV-QKD protocol with a discrete modulation. A Bayesian particle smoother, which is trained using Monte Carlo Markov chain (MCMC) methods, is used for carrier phase estimation at the receiver. As particle smoothing achieves optimum tracking of dynamic variables it is very suitable to investigate the limits of carrier phase estimation in the ultra-low SNR regime. In any CV-QKD protocol, it is absolutely necessary that Alice reveals a fraction of her transmitted symbols via a public classical channel \cite{grosshans2003a}. We demonstrate that these already available symbols can also be used to substantially improve the phase noise mitigation which enables pilotless CV-QKD without any loss of efficiency.
%The investigation in this article focuses on fiber-based systems for CV-QKD. However, the applied methods could also be useful to improve the sensitivity of optical coherent satellite communications.

\section{Results}
\begin{figure*}
	\centering
		\includegraphics{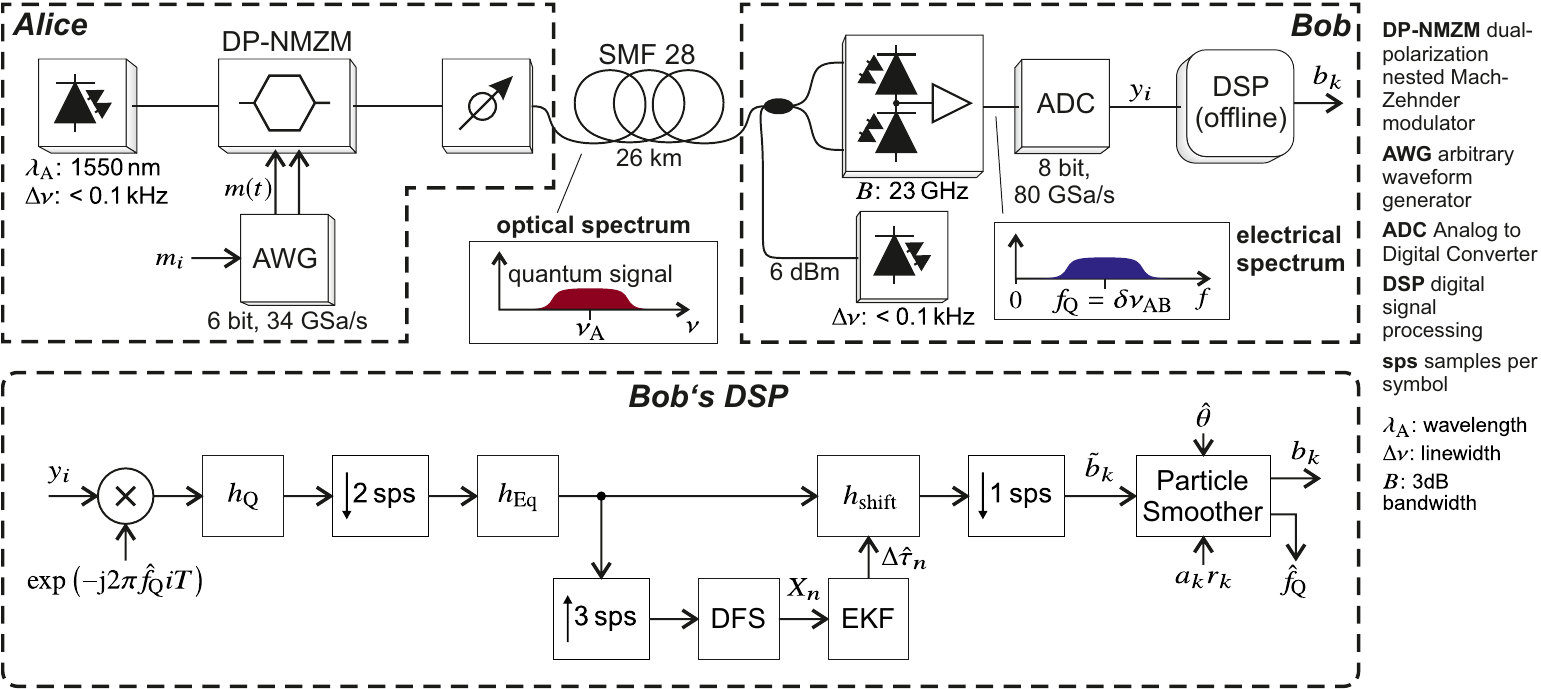}
	\caption{Experimental setup and digital signal processing (DSP) routine for the CV-QKD system without pilot tones. The quantum signal is discrete phase-modulated with a modulation order $M = 2$ or $M = 4$ at a symbol rate of \SI{17}{\giga\baud}. The received and digitized signal is recorded for offline processing. For the DSP routine, the total signal is split into blocks with index $n$. Each block contains $L=\num{6.8e5}$ symbols. Publicly revealing a randomly selected part of Alice's symbols $a_k$ is necessary in CV-QKD \cite{lodewyck2005}. Carrier phase estimation is performed by a particle smoother that takes into account these available symbols.}
	\label{fig:setup_and_DSP}
\end{figure*}
The experimental setup and digital signal processing (DSP) routine is shown in figure \ref{fig:setup_and_DSP}. The quantum signal is discrete phase-modulated in baseband with a modulation order $M=2$ or $M=4$ at a symbol rate of \SI{17}{\giga\baud}. Bob uses a balanced receiver to perform heterodyne detection at an intermediate frequency of $\delta \nu _\mathrm{AB} \approx \SI{10}{\giga\hertz}$. In the DSP routine, Bob first down-converts the signal to baseband and applies the matched filter $h_\mathrm{Q}$. Then, the signal is down-sampled to 2 samples per symbol (sps) and equalized by the FIR-filter $h_\mathrm{Eq}$, which is tuned using the constant-modulus algorithm only once for a high SNR. Timing synchronization is performed by the digital filter and square algorithm \cite{oerder1988} followed by an extended Kalman filter that tracks the argument of the complex Fourier coefficient $X_n$, where $n$ is the block index of the blockwise DSP procedure. After timing correction ($h_\mathrm{shift}$) and down-sampling to one sample per symbol, carrier phase estimation is performed by a particle smoother.
%taking into account the symbols that are revealed by Alice. The revelation of a fraction of Alice's transmitted symbols is necessary in CV-QKD.
In order to find the optimum parameters $\hat{\theta}$ of the state space model, an extended Kalman filter is trained using MCMC methods \cite{sarkka2013}.

The achievable key rate and distance of any CV-QKD system is very sensitive to the excess noise power $\xi_b$ normalized to shot noise units because it could provide information to an eavesdropper. To obtain an accurate excess noise estimate after the quantum communication, receiver noise calibration and quantum signal power estimation are required in CV-QKD.
%For the investigation of our experimental system, the excess noise is estimated in the following way.
Therefore, excess noise estimation is also an important part of our experimental investigation.
%Our system can only be suitable for CV-QKD if it achieves sufficiently low excess noise.
%For that reason, excess noise estimation 
One can write the mean power of the received symbols $b_k$ as
 \begin{gather}
P_b = P_\mathrm{Q} + P_\mathrm{SN} + P_\mathrm{EN} + \xi_b,
\end{gather}  
where $P_\mathrm{Q}$, $P_\mathrm{SN}$, $P_\mathrm{EN}$ and $\xi_b$ are the quantum signal power, the shot noise power, the electrical receiver noise power and the excess noise power in arbitrary units respectively. The excess noise in shot noise units is calculated as
%The achievable key rate and distance of any CV-QKD system is very sensitive to $\xi_b'$ which is the excess noise in shot noise units (\si{\snu}).
 \begin{gather}
\xi_b' = 2\frac{P_b - P_\mathrm{Q} - P_\mathrm{SN} - P_\mathrm{EN}}{P_\mathrm{SN}}.
\end{gather}  
%
 %$P_\mathrm{Q}$, $P_\mathrm{SN}$ and $P_\mathrm{EN}$ have to be estimated to calculate
% \begin{gather}
%\xi_b
%\end{gather}  
%
%.
To calibrate $P_\mathrm{SN}$ and $P_\mathrm{EN}$, the quantum signal is deactivated. The total noise power $P_\mathrm{SN} + P_\mathrm{EN}$ and the electrical noise power $P_\mathrm{EN}$ are calibrated in separate measurements with activated and deactivated LO respectively. This is done after each quantum signal measurement. To estimate $P_\mathrm{Q}$ in our experiments, we evaluate the correlation between $a_k$ and $b_k$.

Generally in CV-QKD, to estimate $P_\mathrm{Q}$ and other important parameters such as the channel transmission and the mutual information shared between Alice and Bob, Alice has to reveal a randomly selected subset of her transmitted symbols $a_k$ after the quantum signal transmission \cite{grosshans2003a}. More precisely, she publicly reveals the vector $a_k \cdot r_k$, where
 \begin{gather}
	r_k =
	\begin{cases}
	1, &  \text{with probability } p_r \\[0pt]
	0, & \text{with probability } 1-p_r.
	\end{cases}
\end{gather}  
That means that for each individual quantum symbol, there is a probability $p_r$ of being revealed after the quantum communication.
%% 5 prozent begründen!
%To evaluate $P_\mathrm{Q}$ in our experiments, the complete sequence $a_k$ is used to obtain an estimate that is accurate regardless of $p_r$.
%Symbol revelation is necessary in CV-QKD not only to evaluate $P_\mathrm{Q}$ but also other important parameters such as the channel transmission and the mutual information shared between Alice and Bob \cite{grosshans2003a}.
As a side effect of symbol revelation, the particle smoother can take advantage of the revealed symbols to improve the accuracy of phase estimation.
Depending on the value of each element $r_k$, the particle smoother adapts its measurement model accordingly. This means that phase estimation can only take place after Alice's symbol revelation. However, as the sequence $\tilde{b}_k$ is already down-sampled to one sample per symbol, it is no additional effort to store the sequence $\tilde{b}_k$ compared to $b_k$. And since the continuous variable $b_k$ must be stored for error correction anyway, which takes place after Alice's symbols have been revealed \cite{grosshans2003}, this approach does not affect the feasibility nor the efficiency of the CV-QKD system in any way.

The noise calibration not only enables the excess noise estimation but also the estimation of the electrical to shot noise ratio $\varepsilon_\mathrm{el}$ and the receiver efficiency $\eta$. These are important characteristics as they not only have an impact on the secret key rate but also on the relation between received optical power $N_\mathrm{B}$ in photons per symbol and the SNR of $b_k$ as
 \begin{gather}
	N_\mathrm{B} = \mathrm{SNR}_{b} (1+\varepsilon_\mathrm{el}) / \eta.
\end{gather}  
In the present setup, $\varepsilon_\mathrm{el} \approx \num{0.70}$ and $\eta \approx \num{0.232}$ which results in $N_\mathrm{B} \approx \num{7.33} \,\mathrm{SNR}_{b}$ corresponding to a penalty for $\mathrm{SNR}_{b}$ of \SI{8.65}{\deci\bel} compared to an ideal heterodyne receiver with $\varepsilon_\mathrm{el}=0$ and $\eta=1$.

To initialize the quantum communication, all symbols of the first signal block are revealed by Alice. This helps the particle smoother to obtain an accurate initial estimate $\hat{f}_\mathrm{Q}$ of the quantum channel frequency $f_\mathrm{Q}$ even at ultra low SNR.
%determine the current quantum channel frequency $f_\mathrm{Q}$.
After the first block, the down-conversion frequency is updated with $\hat{f}_\mathrm{Q}$ and the quantum communication begins.

%The excess noise in shot noise units $\xi_b'=2\xi_b / P_\mathrm{SN}$ is the most important performance parameter of a quantum communication system for CV-QKD.
A residual phase noise in $b_k$ induces excess noise that is proportional to the received quantum signal power \cite{kleis2019}. This means that with decreasing $\mathrm{SNR}_b$ the impact of phase errors on the excess noise is reduced. At ultra low $\mathrm{SNR}_b$, the excess noise can be relatively small, even if the carrier phase estimation fails completely. However, in this case the mutual information between Alice and Bob drops drastically which prevents a successful key generation. Therefore, in order to verify a successful signal demodulation at ultra low $\mathrm{SNR}_b$, we also evaluate the hard decision mutual information $I_\mathrm{AB}$ and compare it to its theoretical value.

%However, at ultra low $\mathrm{SNR}_b$ and a limited number of evaluated symbols, the statistical uncertainty of the estimated excess noise power can become significant compared to the quantum signal power. In that regime, the excess noise is not a reliable indicator of a successful signal demodulation.

%However, the excess noise in shot noise units in the demodulated symbols $\xi_b'$ tends to decrease with $N_\mathrm{B}$ because it originates from signal distortions and imperfections of the demodulation. Additionally, $\hat{\xi}_b'$ is affected by inaccuracies of the receiver noise calibration and there is no theoretical reference to compare it with. That is why, at very low SNR, it is not a reliable indicator of whether the signal could be successfully demodulated or not. In contrast, the mutual information between Alice and Bob $I_\mathrm{AB}$ can be compared to the AWGN channel theory and is therefore more suitable for this purpose.

\begin{figure}
	\centering
		\includegraphics{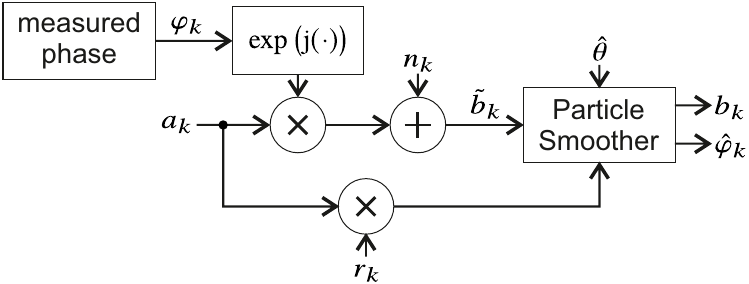}
	\caption{Block diagram for the simulations. The measured phase was obtained using the experimental setup where the SNR of the quantum signal was \SI{7.1}{\deci\bel} and $p_r=1$. The transmission model corresponds to an ideal AWGN channel. For phase estimation, the same particle smoother as in the experiments is used. All other impairments that are present in the experiments such as timing errors, chromatic dispersion, bandwidth limitations and other linear distortions are excluded from the simulations.}
	\label{fig:simulation}
\end{figure}

In addition to the experiments, we carried out simulations to isolate the impact of phase noise and to investigate the phase noise limited performance of our system.
%to gain a deeper insight into the limiting factors of the system.
A block diagram of the simulation model is shown in figure \ref{fig:simulation}. As only additive noise and laser phase noise are included in the simulation model, it does indeed correspond to a phase noise limited system. The phase noise sequence used in the simulations is taken from a measurement with the experimental setup. Thus, the performance of the simulation model corresponds to the best that could be achieved with the laser sources in use.

\begin{figure*}
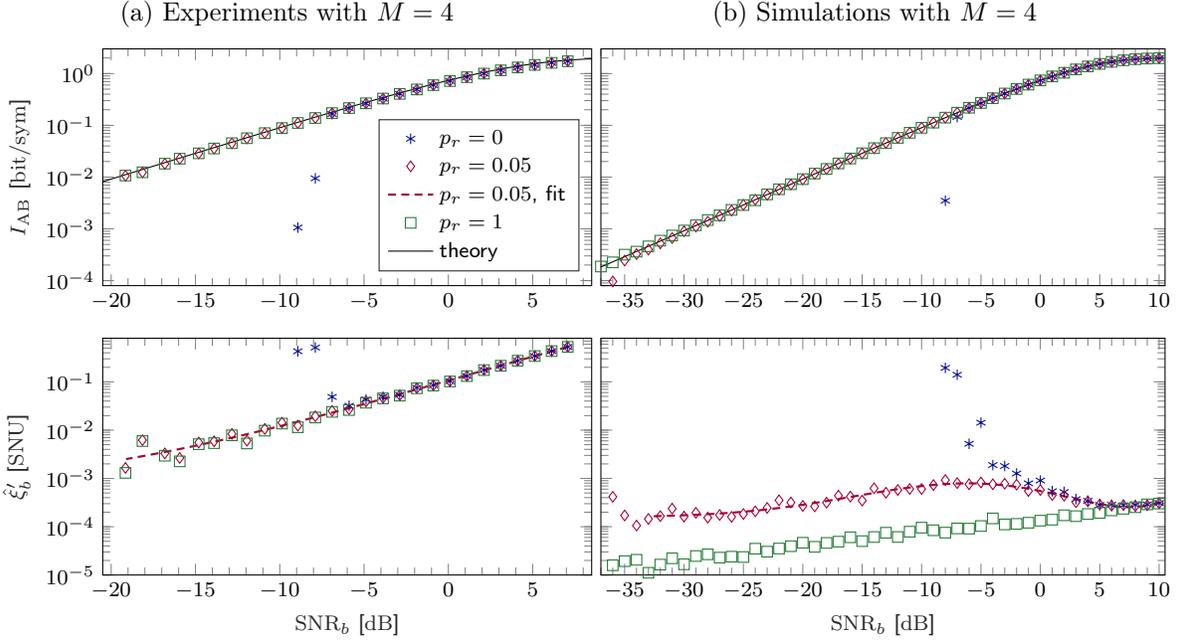

\centering
	\subfloat[Experiments with $M=4$]{
		\centering
		\setlength\figureheight{7cm}
		\setlength\figurewidth{6.5cm}
		\scriptsize
		{\sffamily
		\input{pics/4PSK_exp.tikz}
		}
	}
	\hspace{-0.8cm}
	\subfloat[Simulations with $M=4$]{
		\centering
		\setlength\figureheight{7cm}
		\setlength\figurewidth{7.5cm}
		\scriptsize
		{\sffamily
		\input{pics/4PSK_sim.tikz}
		}	
	}
	\caption{Resulting system performance for $M=4$ in terms of the mutual information between Alice and Bob $I_\mathrm{AB}$ and the excess noise in Bob's received symbols $\xi_b'$. Experimental results are shown on the left hand side (a). Simulation results are shown on the right hand side (b).}
	\label{fig:results_4PSK}
\end{figure*}
\begin{figure*}
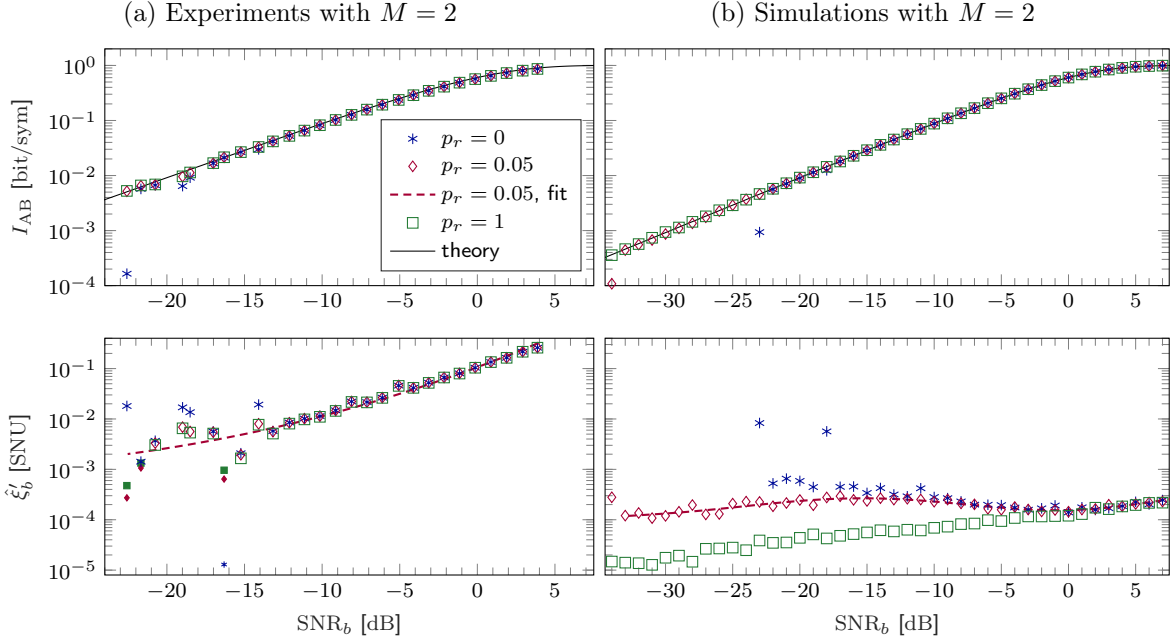

\centering
	\subfloat[Experiments with $M=2$]{
		\centering
		\setlength\figureheight{7cm}
		\setlength\figurewidth{6.5cm}
		{\sffamily
		\scriptsize
		\input{pics/2PSK_exp.tikz}
		}
	}
	\hspace{-0.8cm}
	\subfloat[Simulations with $M=2$]{
		\centering
		\setlength\figureheight{7cm}
		\setlength\figurewidth{7.5cm}
		\scriptsize
		{\sffamily
		\input{pics/2PSK_sim.tikz}
		}	
	}
	\caption{Resulting system performance for $M=2$ in terms of the mutual information between Alice and Bob $I_\mathrm{AB}$ and the excess noise in Bob's received symbols $\xi_b'$. Experimental results are shown on the left hand side (a). Simulation results are shown on the right hand side (b). The small and solid markers indicate negative values.}
	\label{fig:results_2PSK}
\end{figure*}

\section{Discussion}
The experimental and simulation results for $M=4$ and $M=2$ are shown in figures \ref{fig:results_4PSK} and \ref{fig:results_2PSK} respectively.
% Begründung für 5 Prozent?
A typical value of $p_r$ is \num{0.5} \cite{jouguet2013} but in principle it could also be lower. Bob's phase estimation should only use the transmitted symbols that are revealed anyway. Therefore, we investigate our system for a low $p_r$ of \num{0.05}.
%Additionally, we investigate the case $p_r=1$, which represents an upper bound for the performance of phase estimation. Also, we investigate the case $p_r=0$, in 
Additionally, we investigate the cases $p_r=0$ and $p_r=1$. While $p_r=1$ represents an upper bound for the performance of phase estimation, $p_r=0$ shows the achievable performance without symbol revelation. Also, the case $p_r=0$ could be interesting for other fields than CV-QKD where symbol revelation is not possible.

With $p_r = 0$, the receiver sensitvity in the experimental system is $\mathrm{SNR}_{b,\mathrm{min}} = \SI{-6.9}{\deci\bel}$. This is in very good agreement with the simulation, where the signal demodulation completely fails at an $\mathrm{SNR}_b$ of \SI{-8}{\deci\bel}. Thus, despite the narrow linewidth fiber lasers, the receiver sensitivity is limited by phase noise.
In terms of received optical power, at least \SI{1.5}{\photon\per\symbol} would be required.
%In the given experimental setup, this corresponds to a sensitivity of $N_\mathrm{B,min} \approx \SI{1.5}{\photon\per\symbol}$.
This is too high for the discrete phase modulation scheme with $M=4$, where the transmitted optical power should usually be lower than \SI{0.5}{\photon\per\symbol} \cite{leverrier2011}. However, even a small revelation probability of $p_r=\num{0.05}$ is sufficient to enable successful signal demodulation down to at least an $\mathrm{SNR}_b$ of \SI{-19.1}{\deci\bel}. In the experiment, the SNR was not decreased further because the receiver calibration was not accurate enough to allow for a reliable evaluation of $\mathrm{SNR}_b$ and the excess noise in that regime.
%uncertainty induced by the receiver calibration would become very high.
The experimentally confirmed receiver sensitivity of $\mathrm{SNR}_{b,\mathrm{min}} < \SI{-19.1}{\deci\bel}$ is sufficient for CV-QKD over \SI{26}{\kilo\meter} and more.
In the simulation results, it can be seen that the carrier phase estimation can be successful down to an $\mathrm{SNR}_b$ of \SI{-35}{\deci\bel}. However, there is already an increased probability of failure above \SI{-35}{\deci\bel} as can be seen in the simulations for $M=2$ shown in figure \ref{fig:results_2PSK}, where the minimum $\mathrm{SNR}_b$ was $\SI{-33}{\deci\bel}$. Therefore, we take this higher value as the receiver sensitivity.

%The excess noise in the simulations is only induced by phase noise.
The simulations show the phase noise limited performance.
At an $\mathrm{SNR}_b$ of \SI{10}{\deci\bel}, there is no difference in the resulting excess noise between different probabilities of revelation. Apparently, in that case the uncertainty about the carrier phase is not increased by the discrete phase modulation. This is the SNR regime of classical communications where symbol error rates are low and the modulation could also be canceled effectively before carrier phase estimation. In the quantum regime of $\mathrm{SNR}_b<\SI{0}{\deci\bel}$, the modulation clearly affects the carrier phase estimation and hence the excess noise. With $p_r=1$, $\xi_b'$ exhibits a constant slope of about \SI{3}{\deci\bel} per decade. This is the result of a decreasing quantum signal power decreasing the excess noise while at the same time decreasing the SNR increases the carrier phase uncertainty which induces additional excess noise. With $p_r=0$, in the regime of $\mathrm{SNR}_b<\SI{5}{\deci\bel}$, there is a net increase of $\xi_b'$ with decreasing $\mathrm{SNR}_b$ until the demodulation fails completely. Interestingly, for $p_r = 0.05$, there is not much value of the revelation around $\mathrm{SNR}_b \approx \SI{0}{\deci\bel}$. In this regime, the unrevealed symbols still provide significant information about the carrier phase. But in the regime of $\mathrm{SNR}_b < \SI{-8}{\deci\bel}$, the slope of $\xi_b'$ approaches the one of the case $p_r=1$. In this regime, the revealed symbols provide more information about the carrier phase than the unrevealed ones and the difference between $p_r=1$ and $p_r=0.05$ can be interpreted as a difference of the effective symbol rate. Due to this, there is a relatively constant excess noise penalty of about a factor of \num{9} in the $\mathrm{SNR}_b$ regime between \SI{-10}{\deci\bel} and \SI{-26}{\deci\bel}. Below an $\mathrm{SNR}_b$ of $\SI{-26}{\deci\bel}$, the slope is flattened.

Looking at the experimental results for the excess noise, we observe a much higher level than in the simulations. Also, there is a proportional relation between $\hat{\xi}_b'$ and $\mathrm{SNR}_b$ that slightly flattens below an $\mathrm{SNR}_b$ of \SI{-7}{\deci\bel}. This confirms that the excess noise mainly originates from signal distortions. The signal to distortion ratio can be quantified for $\mathrm{SNR}_b=\SI{7.1}{\deci\bel}$ as $P_\mathrm{Q} / \xi_b \approx \num{31.82}$, which is relatively low and could be improved by optimizing the equalization concept.

For $M=2$, the results are shown in figure \ref{fig:results_2PSK}. The main difference to $M=4$ is that the receiver sensitivity for the case $p_r=0$ is much lower. In the simulations it was $\mathrm{SNR}_{b,\mathrm{min}} = \SI{-17}{\deci\bel}$ and in the experiment $\mathrm{SNR}_{b,\mathrm{min}} = \SI{-13.2}{\deci\bel}$ compared to the \SI{-6.9}{\deci\bel} for $M=4$. The difference between experiment and simulations seems quite large. However, there is still a significant probability of successful demodulation below these values in the experiments and simulations. This indicates that the probability of failure increases less steep with decreasing $\mathrm{SNR}_b$ as in the case of $M=4$. With $p_r=\num{0.05}$, the transition to relying only on the revealed symbols for phase estimation is much wider for $M=2$ compared to $M=4$. Even at $\mathrm{SNR}_b=\SI{-30}{\deci\bel}$, the excess noise is slightly lower with $M=2$ compared to $M=4$. However, there is no significant difference between these two cases in terms of experimental excess noise which is dominated by other signal distortions.

\begin{figure}%
	\centering
	\setlength\figureheight{4.5cm}
	\setlength\figurewidth{10cm}
	\scriptsize
	{\sffamily
	\input{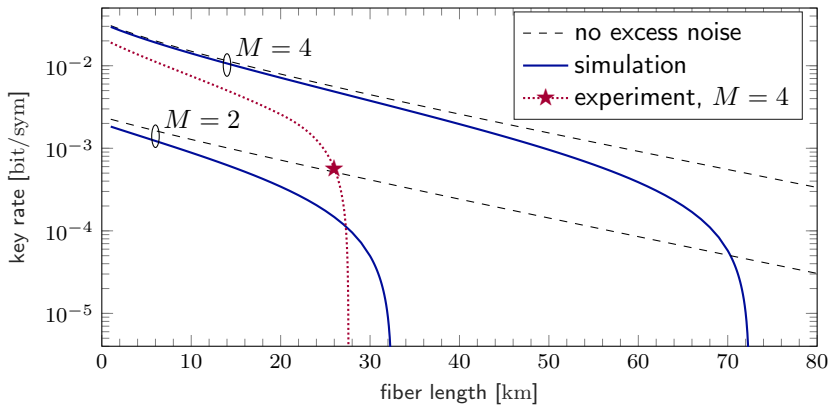}
	}
	\caption{Achievable key rates for $p_r=0.05$, taking into account the polynomial fits for the excess noise in figure \ref{fig:results_4PSK} as well as the experimental parameters $\hat{\eta}$, $\hat{\varepsilon}_\mathrm{el}$ and $\mathrm{SNR}_{b,\mathrm{min}}$. A reconciliation efficiency of \SI{95}{\percent} and a fiber loss of \SI{0.2}{\deci\bel\per\kilo\meter} was assumed. The launch power was optimized based on the security analysis in \cite{leverrier2011}.}
	\label{fig:keyrates}
\end{figure}
In order to address the question of whether or not the performance of the investigated system with a real LO and without pilots is sufficient for CV-QKD, we calculated asymptotic secret key rates. 
The calculations are based on the well established security proofs \cite{leverrier2010} and \cite{leverrier2011}.
These assume a linear Gaussian channel and are therefore not fully general. Recently, a security proof that does not incorporate these assumptions has been presented \cite{ghorai2019}. However, the question of how to get rid of the linear Gaussian assumption is still under discussion. 

We optimized the transmitted optical power with respect to a maximum key rate.
For this, we used the polynomial fits for the excess noise in the case of $p_r=\num{0.05}$ that are plotted in the figures \ref{fig:results_4PSK} and \ref{fig:results_2PSK}. An additional constraint of the optimization was that $\mathrm{SNR}_b$ is not allowed to be lower than in the experiments and simulations respectively and also not allowed to be lower than the receiver sensitivity. The measured receiver efficiency $\hat{\eta}$, the electronic noise ratio $\hat{\varepsilon}_\mathrm{el}$ as well as the fiber length are also included in the secret key rate calculations. The reconciliation efficiency is assumed to be \SI{95}{\percent}.

The resulting secret key rates for $p_r=\num{0.05}$ are shown in figure \ref{fig:keyrates}. As a reference, the case of zero excess noise is also plotted.
%Without any phase noise the achievable distance would not be limited. This case is plotted as a reference. 
The simulation corresponds to the case of the excess noise being dominated by carrier phase uncertainty. In this case, distances of up to \SI{72}{\kilo\meter} could be achieved with $M=4$ and \SI{32}{\kilo\meter} with $M=2$. This shows that laser phase noise is not an insurmountable obstacle for designing a CV-QKD system with a real LO and without pilots. Even for the experimental system, where the excess noise is much higher due to other signal distortions, CV-QKD could be performed successfully with $M=4$ at a key rate of \SI{5.7e-4}{\bit\per\symbol} over the experimental distance of \SI{26}{\kilo\meter}. With $M=2$, the experimental performance is not sufficient for CV-QKD. The worse performance of $M=2$ is due to stricter requirements regarding the transmitted power and excess noise \cite{leverrier2011}.
The experimental key rate can be improved towards the simulation by reducing signal distortions in the experiment. This can be achieved by improving the equalization concept of the system. To achieve the long distances that are possible as shown by the simulations, the receiver calibration should be improved. We expect that this can be achieved by reducing the time difference between calibration and quantum signal transmission in order to mitigate fluctuations of the receiver noise and LO power.
If the system is optimized such that it is limited by phase noise, the parameter $p_r$ can be increased which directly reduces the excess noise for even higher key rates and longer distances. Also, increasing $p_r$ can enable the use of lasers with stronger phase noise while keeping the achievable distance.

To conclude, we investigated a fiber-based quantum communication system for CV-QKD that employs a real LO and works without any pilot tones. 
As Bob's clock and LO are free running, he
relies only on the modulated quantum signal itself to perform carrier phase estimation and timing recovery.
Except for the fact that the signal is attenuated before it is transmitted, the physical implementation is identical to classical coherent systems.
For the first time it could be demonstrated experimentally, that such a system can be feasible for CV-QKD. An important factor to achieve this is the particle smoother that is used for carrier phase estimation, which is optimized using Monte Carlo Markov chain methods. Based on the experimental results, the achievable key rate over a distance of \SI{26}{\kilo\meter} is \SI{5.7e-4}{\bit\per\symbol} corresponding to \SI{9.2}{\mega\bit\per\second}. The simulation results indicate that this performance can even be largely improved. Possible ways to achieve this is by reducing other signal distortions than phase noise and by improving the receiver calibration.
The achieved results are an important milestone for CV-QKD. Also, the achieved receiver sensitivities without symbol revelation ($p_r=0$) of \SI{-6.9}{\deci\bel} for $M=4$ and \SI{-13.2}{\deci\bel} for $M=2$ have, to our knowledge, never been reached before. Therefore, the novel techniques are also beneficial in other coherent systems where the SNR can be extremely low, such as optical satellite communications.
%While this is an important milestone for CV-QKD, the used techniques to achieve, without symbol revelation ($p_r=0$), receiver sensitivities of \SI{-6.9}{\deci\bel} for $M=4$ and \SI{-13.2}{\deci\bel} for $M=2$ could also be useful for other coherent systems where the SNR can be extremely low, such as optical satellite communications.

\section{Methods}
\subsection{Experimental Details}
The transmitter laser and LO are continuous wave DFB fiber lasers of type NKT Koheras E15. Only one polarization of the dual polarization modulator is used, the orthogonal one is biased to the zero transmission point. The baseband quantum signal $m_i$ contains Alice’s symbols $a_k$ that are pulse-shaped
by the root raised cosine filter $h_\mathrm{Q}$ with a roll-off factor of \num{0.1} and a bandwidth of \SI{17}{\giga\hertz}.

For each experimental scenario, defined by the transmitted optical power and modulation order, Bob's receiver is calibrated after quantum signal transmission. For this, the quantum signal is deactivated to record the total receiver noise. After that, the LO is also deactivated to record only the electrical receiver noise. The noise signals undergo the same DSP routine as the quantum signal but with deactivated timing recovery and phase estimation because these methods do not alter the evaluated mean power of the received noise sequence. The signals are recorded with a time difference of about \SI{3}{\second}. For each quantum and noise signal, a total number of 100 blocks are evaluated corresponding to \num{6.55e6} symbols $b_k$.  This does not include the first signal block that is used to initialize $\hat{f}_\mathrm{Q}$. The transmitted optical power in the initialization block is the same as for the subsequent quantum communication.

\subsection{Simulation Details}
In the simulations, the number of evaluated quantum symbols is the same as in the experiment.
The true phase $\varphi_k$ is directly accessible. Excess noise due to inaccuracies in the estimated phase $\hat{\varphi_k}$ can be calculated as \cite{kleis2019}
 \begin{gather}
\xi_b' = 2 \frac{P_\mathrm{Q}}{P_\mathrm{SN}} \langle \sin^2 ( \varphi_k - \hat{\varphi_k} ) \rangle, \\
P_\mathrm{Q} = \langle |a_k|^2 \rangle.
\end{gather}  
In the simulations, only the total noise power $P_\mathrm{TN}$ is specified. To calculate $\xi_b'$, one must assume
a specific electrical to shot noise ratio $\varepsilon_\mathrm{el}$ to calculate
 \begin{gather}
\xi_b' = 2\, \mathrm{SNR}_b \,(1+\varepsilon_\mathrm{el})\, \langle \sin^2 ( \varphi_k - \hat{\varphi_k} ) \rangle.
\end{gather}  
Here, $\varepsilon_\mathrm{el}$ is set to \num{0.7} which corresponds to the value that was measured in the experiments.

\subsection{Particle Smoother}
The particle smoother is a Bayesian method that requires a measurement model and a dynamic model formulated as probability densities. A detailed general description of particle smoothing can be found in \cite{sarkka2013}.
The transmitted symbols are $M$-ary phase modulated, meaning that
 \begin{gather}
a_k \in \left\{ a^{(1)}, \ldots ,a^{(M)} \left| a^{(i)} = \mathrm{e} ^{\,\mathrm{j} 2 \pi \frac{i}{M}} \right. \right\}.
\end{gather}  
In an AWGN channel, Bob's measurement signal at the input of the particle smoother can be written as
 \begin{gather}
\tilde{b}_k = a_k \mathrm{e} ^{\,\mathrm{j}\varphi_k} + n_k,
\end{gather}  
where $n_k$ is a complex white Gaussian noise sequence with mean power $P_\mathrm{TN}$ and $\varphi_k$ is the dynamic laser phase. The normalization of Bob's signal such that it matches $|a_k|=1$ is performed using the noise power that is known from receiver calibration. Thus, for the case $r_k=0$, the measurement probability density can be written as

\begin{align}
p\bigl(\tilde{b}_k|\varphi_k\bigr) &= \frac{1}{M}\sum\limits_{i=1}^{M} p\bigl(\tilde{b}_k|a^{(i)},\varphi_k\bigr) \\
&= \frac{1}{M}\sum\limits_{i=1}^{M} \mathcal{N} \bigl(a^{(i)} \mathrm{e} ^{\,\mathrm{j}\varphi_k} , P_\mathrm{TN} / 2 \cdot I_2 \bigr).
\end{align}

Here, $\mathcal{N} (\mu,\Sigma)$ is a Gaussian probability density with mean $\mu$ and covariance matrix $\Sigma$. $I_2$ denotes the 2-dimensional identity matrix. If $r_k=1$, the transmitted symbol $a_k$ is known to Bob and the measurement probability density reduces to $p\bigl(\tilde{b}_k|\varphi_k\bigr) = \mathcal{N} \bigl(a_k \mathrm{e} ^{\,\mathrm{j}\varphi_k} , P_\mathrm{TN} / 2 \cdot I_2 \bigr)$. There are no unknown parameters in the measurement model.

The state space model $p(\mathbf{x}_{k}|\mathbf{x}_{k-1})$ describes the dynamics of the variable $\mathbf{x}_k$. We define it as

\begin{align}
p(\mathbf{x}_{k}|\mathbf{x}_{k-1}) &=  p(\varphi_k|\varphi_{k-1},\Omega_{k-1}) \cdot p(\Omega_k|\Omega_{k-1}) \\
&=   \mathcal{N} ( \varphi_{k-1} + \Omega_{k-1}, \sigma_\varphi ^2) \cdot \mathcal{N} ( \Omega_{k-1}, \sigma_\Omega ^2).
\label{eq:state_model}
\end{align}

The variable $\Omega_k$ is a normalized frequency that models a drift of the differential laser frequency as a random walk with variance $\sigma^2_\Omega$ \cite{piels2015}. Additionally, $\varphi_k$ is affected by a random walk with variance $\sigma_\varphi^2$. Based on the stated measurement and dynamic model, the particle smoother is implemented as a bootstrap filter with $N=\num{200}$ particles in combination with a backward-simulation particle smoother with $\num{10}$ trajectories. The resampling condition of the bootstrap filter is $N_\mathrm{eff}<N/5$, where $N_\mathrm{eff}$ is the effective number of particles.

\subsection{Bayesian Parameter Optimization}
The particle smoother is capable of performing optimum phase estimation, given that the state space model is an accurate representation of the measurement and laser dynamics. Thus, optimizing the set of unknown parameters $\theta = [\sigma^2_\Omega,\sigma_\varphi ^2]$ is essential. For this, we used Monte Carlo Markov chain methods which are based on minimizing the energy function $\Phi_K(\theta)$. The energy function has the property $\mathrm{e} ^{-\Phi_K(\theta)} \propto p(\theta|\tilde{b}_{1:K})$. Therefore, minimizing it leads to the parameters $\hat{\theta}$ that provide the most accurate description of the true phase dynamics. We performed the minimization based on a received signal $\tilde{b}_k$ with $\mathrm{SNR}_b=\SI{11.5}{\deci\bel}$ and $p_r=1$ using an extended Kalman filter to calculate $\Phi_K(\theta)$ for $K=\num{8.1e6}$. The extended Kalman filter is based on the same state space model as the particle smoother. The minimum was found using the simplex search method \cite{lagarias1998}. The resulting optimized parameters are $\hat{\sigma}_\Omega^2=\SI{1.66e-16}{\radian\squared}$ and $\hat{\sigma}_\varphi^2=\SI{6.36e-9}{\radian\squared}$.

\subsection{Timing Recovery}
For signal processing, the total signal is split into consecutive blocks of length $L=\num{6.8e5}$ symbols with index $n$.
The digital filter and square algorithm for timing recovery is implemented as described in \cite{oerder1988} and calculates the complex Fourier coefficient $X_n$ using one complete block. From the argument of $X_n$, the current timing offset can be obtained. The SNR of $X_n$ scales approximately as $\mathrm{SNR}_X \propto \mathrm{SNR}_b^2 L$ in the regime of $\mathrm{SNR}_b \ll 1$. Thus, the accuracy of timing recovery can be improved by increasing $L$. However, the timing experiences a relatively constant drift of $0.032$ symbol periods per block due to a clock offset between Alice and Bob of \SI{4.7e-8}{\hertz\per\hertz} which means that the block length cannot be arbitrarily increased. At the given block length, pulse shape and at an $\mathrm{SNR}_b$ of \SI{-20}{\deci\bel}, the resulting $\mathrm{SNR}_X$ is only \SI{-0.75}{\deci\bel} which is too low to obtain sufficiently low excess noise.
Therefore, our approach is to track the argument $X_n$ using an extended Kalman filter. As this is very similar to the problem of carrier phase tracking, the same state space model is used.

\subsection{Estimation of Mutual Information and its Theoretical Value}
For mutual information estimation, Bob performs a hard decision on his symbols, meaning that each $b_k$ is mapped to the symbol $a^{(i)}$ of the discrete alphabet with the smallest euclidean distance. Bob's symbols after this mapping are denoted as $\hat{a}_k$. Based on the measured probabilities $p(\hat{a}_k=a^{(i)},a_k=a^{(j)})$ we calculate the mutual information, where all transmitted and received symbols are taken into account.

Assuming an ideal AWGN channel, the received symbols can be written as
% \begin{gather}
$b_k = \sqrt{\mathrm{SNR}_b} \cdot a_k + n_k$,
%\end{gather}  
where $a_k=\e^{\mathrm{j}\alpha_k}$ is Alice's transmitted symbol and $n_k$ is a white Gaussian noise sequence with a mean power of 1. The probability density of the phase $\beta_k=\arg(b_k)$ is then
 \begin{gather}
p(\beta_k) = \frac{1}{2\pi} \e^{-\mathrm{SNR}_b} + \frac{\sqrt{\mathrm{SNR}_b}}{2\sqrt{\pi}} \cos(\alpha_k-\beta_k) \e^{-\mathrm{SNR} \sin^2(\alpha_k-\beta_k)} \erfc\left(-\sqrt{\mathrm{SNR}_b} \cos(\alpha_k-\beta_k)\right).
\end{gather}  
By numerical integration of $p(\beta_k)$ over each decision region and for each transmitted symbol, we obtain the theoretical probabilities $p(\hat{a}_k=a^{(i)},a_k=a^{(j)})$ that are required to calculate the theoretical value of $I_\mathrm{AB}$.

\subsection{Estimation of Quantum Signal Power}
For excess noise estimation, the quantum signal power $P_\mathrm{Q}$ must be estimated. In our experiments, this is done by calculating 
 \begin{gather}
P_\mathrm{Q}=\left| \frac{1}{K} \sum\limits _{k=1} ^K a_k b^*_k \right|^2,
\end{gather}  
where $K$ is the total number of evaluated symbols. For the estimation of $P_\mathrm{Q}$, all transmitted and received symbols are used to obtain an estimate that is accurate regardless of $p_r$.
\end{document}